\documentstyle[11pt,psfig]{article}
\title{\bf Quantum Monte Carlo calculations of the structural
properties and the B1-B2 phase transition of MgO }
\author{D. Alf\`{e}$^{1,2}$, M. Alfredsson$^1$, J. Brodholt$^1$ and
M. J. Gillan$^2$
\smallskip \\ $^1$Department of Earth Sciences, University College
London \\ Gower Street, London WC1E~6BT, UK
\smallskip \\ $^2$Department of Physics and Astronomy, University
College London \\ Gower Street, London WC1E~6BT, UK
\bigskip \\ M. D. Towler and R. J. Needs
\smallskip \\ Theory of Condensed Matter Group, Cavendish Laboratory, 
University of Cambridge \\ Cambridge CB3~0HE, UK}

\begin{document}

\maketitle

\begin{abstract}
We report diffusion Monte Carlo (DMC) calculations on MgO in the
rock-salt and CsCl structures. The calculations are based on
Hartree-Fock pseudopotentials, with the single-particle orbitals
entering the correlated wave function being represented by a
systematically convergeable cubic-spline basis. Systematic tests are
presented on system-size errors using periodically repeating cells of
up to over 600 atoms. The equilibrium lattice parameter of the
rock-salt structure obtained within DMC is almost identical to the
Hartree-Fock result, which is close to the experimental value. The DMC
result for the bulk modulus is also in good agreement with the
experimental value. The B1-B2 transition pressure (between the
rock-salt and CsCl structures) is predicted to be just below 600 GPa,
which is beyond the experimentally accessible range, in accord with
other predictions based on Hartree-Fock and density functional
theories.
\end{abstract}

\section{Introduction}

The quantum Monte Carlo (QMC) technique is becoming an increasingly
important tool in the study of condensed
matter~\cite{foulkes01}. Competitive in accuracy with high-level
quantum chemistry methods, it has the enormous advantage of being
practicable for large systems containing hundreds of atoms. The power
of QMC in overcoming the deficiencies of density functional theory
(DFT) has been amply demonstrated by recent applications, including
the energetics of point defects in silicon~\cite{leung99} and
carbon~\cite{hood03}, the reconstruction of the Si (001)
surface~\cite{healy01} and its interaction with
H$_2$~\cite{filippi02}, and the calculation of optical excitation
energies~\cite{williamson02}. Nevertheless, the classes of materials
to which QMC has been applied have so far been rather limited.  Oxide
materials are likely to be a very fruitful field for the application
of QMC, but in exploring this field it is clearly important to study
the capabilities of the techniques for the simplest possible
oxides. We present here QMC calculations on MgO, focusing on its
elementary bulk properties, including the equilibrium lattice
parameter of the rock-salt structure, the stable structure under
ambient conditions, and the pressure of the B1-B2 transition between
the rock-salt and CsCl structures.

Two types of QMC are relevant here. In the first, known as variational
Monte Carlo (VMC), a trial many-electron wave function is constructed
as the product of a Slater determinant of single-electron orbitals and
the so-called Jastrow correlation factor. The latter is parameterised,
and the parameter values are obtained using a stochastic optimisation
procedure.  Since VMC by itself is not usually accurate enough, the
optimised many-electron wave function produced by VMC is then used in
diffusion Monte Carlo (DMC)~\cite{foulkes01,ceperley80}, which
improves the ground-state estimate by performing an evolution in
imaginary time. In principle, the ground-state energy would be exact,
but to overcome the fermion sign problem we use the standard
``fixed-node approximation''~\cite{anderson76}.  In practice, only the
valence electrons are treated explicitly, the interactions between the
valence and core electrons being represented by pseudopotentials. This
introduces additional approximations, including the ``pseudopotential
locality approximation''. The calculations are performed on
periodically repeated cells, and system size errors need to be
carefully treated. However, in many cases, the overall errors within
QMC can be made much smaller than those within DFT, and QMC has
already been important in revealing, quantifying and overcoming DFT
errors in such quantities as defect formation energies, surface
reaction energies and energy barriers~\cite{leung99,hood03,filippi02}.

The three main purposes of this work are: first, to establish the
technical feasibility of performing QMC on MgO; second, to study
differences between DFT and QMC predictions for the properties of bulk
MgO; and third, to prepare the way for QMC work on more challenging
oxides. As one of the simplest oxides, MgO has often been used as a
paradigm for testing theoretical techniques. For QMC, the issue of
technical feasibility is a non-trivial one, since the computing effort
required to obtain accurate results with DMC depends heavily on the
ability of VMC to deliver wave functions which are sufficiently close
to the exact ground-state wave function. If sufficiently accurate
trial wave functions cannot be obtained, the DMC calculations may even
become unstable.  The electronic simplicity of MgO is expected to ease
the task of finding good trial wave functions.

The concerns of this work are not purely technical. The
pressure-induced transition from the rock-salt to CsCl structure in
MgO has been much studied because of its geological interest (see,
e.g.,~\cite{oganov03,oganov03prb} and references therein).  We also
expect that the present work will provide the basis for applying QMC
to several important and controversial questions related to ionic
materials such as MgO, including the adsorption and dissociation of
molecules on their surfaces (see, e.g., Ref.~\cite{lindan2003}), the
self-trapping of hole centres and their trapping at defects (see,
e.g., Ref.~\cite{gavartin2003}), and the conflict between theory and
experiment for the slope, $dT/dp$, of their melting
curves~\cite{zerr1994,cohen1994,vocadlo1996,strachan99}.  Beyond this,
we hope that the experience gained here will help to prepare the way
for the application of QMC to transition-metal oxides such FeO, where
electron correlation is highly non-trivial.  DMC studies of
NiO~\cite{needs_2003} and MnO~\cite{lee_2004} have already shown the
feasibility of calculations on these materials, but those studies did
not include energy-volume curves, and the unit cells used were not
large enough to give the accuracy required here.

In the following Section we summarise the QMC techniques used here. In
Sec.~3, we present tests on the magnitude of various errors, including
system-size errors, and we report our results for the total energy as
a function of volume for the rock-salt and CsCl structures, and the
transition pressure between the two. Discussion, prospects for future
work, and conclusions are presented in Sec.~4.

\section{Techniques}

Detailed descriptions of VMC and DMC have already been
reported~\cite{foulkes01}, so here we only outline rather briefly the
main features of the present work. All the QMC calculations presented
here were performed using the {\sc casino} code, the technical details
of which are given in Ref.~\cite{needs04}.

Our trial wave functions are of the Slater-Jastrow type:
\begin{equation}
\label{eq:SJ}
\Psi_T ( {\bf R} ) = D^\uparrow D^\downarrow e^{J} \; ,
\end{equation}  
where $D^\uparrow$ and $D^\downarrow$ are Slater determinants of up-
and down-spin single-electron orbitals, and $e^J$ is the so-called
Jastrow factor, describing the correlation between the electrons.  We
use single-electron orbitals obtained from DFT calculations. These
single-particle orbitals fix the nodal surface (the surface in
configuration space on which the wave function vanishes and across
which it changes sign). Within this ``fixed-node approximation'' DMC
gives a variational upper bound to the ground state energy, rather
than the exact ground state energy.
However, because of its large band-gap, we expect that a single Slater
determinant will give a good description of the nodal surface of
MgO. The function $J$ appearing in the Jastrow factor is a sum of
parametrized one-body and two-body terms~\cite{williamson96}, the
latter being designed to satisfy the cusp conditions. The free
parameters in $J$ are determined by requiring that the variance of the
local energy in VMC be as small as possible~\cite{umrigar88,kent99b}.

In the present work, the many-body wave function represents explicitly
only valence electrons, whose interactions with the ionic cores are
represented by pseudopotentials. We used pseudopotentials generated
within Hartree-Fock (HF) theory, but including scalar relativistic
effects~\cite{trail_2004}.  There is evidence to show that HF theory
provides better pseudopotentials for use within QMC than
DFT~\cite{greeff98}. The core radii of our pseudopotentials are
$r({\rm O2s}) = 0.423$~\AA, $r({\rm O2p}) = 0.397$~\AA, 
$r({\rm O3d}) = 0.524$~\AA, 
$r({\rm Mg3s}) = r({\rm Mg3p}) = r({\rm Mg3d}) = 1.259$~\AA.

The single-particle orbitals entering the Slater determinants of
Eq.~(\ref{eq:SJ}) are the most important component of the wave
function.  Filippi and Fahy~\cite{filippi00} have developed a method
for optimising orbitals within VMC, which achieved an energy reduction
in diamond from optimising LDA orbitals of 0.040(16) eV per atom.
Kent \textit{et al.}~\cite{kent98} found that in bulk silicon using
LDA orbitals in a VMC calculation gave an energy 0.024(4) eV per atom
lower than HF orbitals.  These energy changes would be significantly
reduced in DMC and it appears that HF and LDA orbitals are sufficient
for these systems.  In this work we have used LDA orbitals obtained
using the plane-wave pseudopotential DFT code {\sc
pwscf}~\cite{PWSCF}.

The basis set used to represent the single-particle orbitals in the
QMC calculations themselves is not plane-waves, which become very
inefficient for large systems, because the computation cost of
evaluating an orbital is proportional to the system size. Instead, we
use a B-spline basis, also known as blip functions, consisting of
piecewise continuous localised cubic spline functions centred on a
regular grid of points. A detailed explanation of blip functions, and
their great advantages for QMC calculations has been reported
elsewhere~\cite{alfe04}. The blip basis is closely related to a
plane-wave basis, and for a plane-wave cut-off energy $E_{\rm cut}=
\hbar^2 k_{\rm cut}^2/2m$ ($m$ is the electron mass), there is a
natural choice of blip-grid spacing $a$ given by $a=\pi/k_{\rm cut}$. 
In the same way that plane-wave convergence is achieved by
increasing $k_{\rm cut}$, blip convergence is achieved by reducing
$a$.  Because of the relationship between plane-waves and blips, it is
straightforward to transform the plane-wave coefficients from the {\sc
pwscf} calculations into the blip coefficients needed for the QMC
calculations, as explained in more detail in our earlier
paper~\cite{alfe04}.

For QMC calculations on perfect crystals, there is a useful device
which allows a considerable saving of memory. Instead of constructing
single-particle orbitals at a given {\bf k}-point (e.g., the $\Gamma$
point) for the large repeating cell, we construct them for the
primitive cell at the corresponding set of {\bf k}-points. The plane
wave coefficients from this calculation are then converted to blip
coefficients on points of the blip-grid within the primitive cell. At
run-time, a simple conversion allows these stored coefficients to be
used to calculate the required values of the single-particle orbitals
at any point in the large repeating cell. The key point here is that
it is unnecessary to store blip coefficients at grid points covering
the entire large repeating cell.

An important source of error in QMC calculations using periodic
boundary conditions is the limited size of the repeating cell, and the
convergence of the QMC energy with respect to the size of the
simulation cell must be carefully investigated. To improve this
convergence, we follow the common practice~\cite{williamson98} of
correcting for this error by using separate DFT calculations: we add
to the DMC energies the difference $\Delta E_{\Gamma \rightarrow \bf
k}$ between the DFT-LDA energy calculated with a very large set of
{\bf k}-points and the DFT-LDA energy calculated using the same
sampling as in the DMC calculation. The question of correcting for
finite size errors in the Coulomb energy has been addressed in recent
papers~\cite{fraser96,williamson97,kent99}, and a method known as the
model periodic Coulomb (MPC) interaction has been developed. The
finite size error in the Ewald interaction energy arises from the
exchange-correlation energy, which can be written as the interaction
of the electrons with their exchange-correlation holes.  The
interaction with the hole should have the standard $1/r$ form, but
within periodic boundary conditions this must be replaced by a
periodic interaction.  The MPC interaction maintains the correct Ewald
interaction for evaluating the Hartree energy while for the
exchange-correlation energy a periodically repeated potential based on
the $1/r$ form is used.  This significantly reduces the finite size
errors in the interaction energy, although effects due to the
squeezing of the exchange-correlation hole into a finite cell still
remain.

We also mention two other technical points relating to size
effects. DMC calculations require the use of real trial wave
functions. However, these can be constructed using single-particle
orbitals obtained either from calculations at the $\Gamma$-point or,
in general, {\bf k}-points which correspond to one half of a
reciprocal lattice vector of the simulation cell~\cite{maezono03}. The
difference between QMC energies obtained in these two ways can be used
as an indication of the system size errors. The other point is that a
given physical crystal structure can be treated using large repeating
cells associated with different Bravais lattices. Since MgO in both
the rock-salt and CsCl structures has cubic symmetry, it is
most natural to use
Bravais lattices for the repeating cell having
simple-cubic (sc), body-centred-cubic (bcc) or
face-centred-cubic (fcc) symmetries. 
We expect that the fcc repeating geometry will give the
best convergence with respect to system size, and we shall present
results which illustrate this effect.
 
The number of walkers in DMC simulations is governed by a population
control algorithm, which has the purpose of maintaining this number
roughly constant.  In order to minimise statistical bias in the total
energy, the calculations need to be run with a large population of
walkers. For our DMC calculations we have used a target population of
640 walkers, which also makes it efficient to run on massively
parallel machines, with parallelism achieved by distributing walkers
across processors.  For the imaginary time evolution of the walkers we
found that a time step of 0.005 a.u.\ gave time step errors in the DMC
energy of less than 10 meV/atom.

\section{Results}

\subsection{Technical tests}
\label{subsec:technical_tests}
\begin{figure}
\psfig{figure=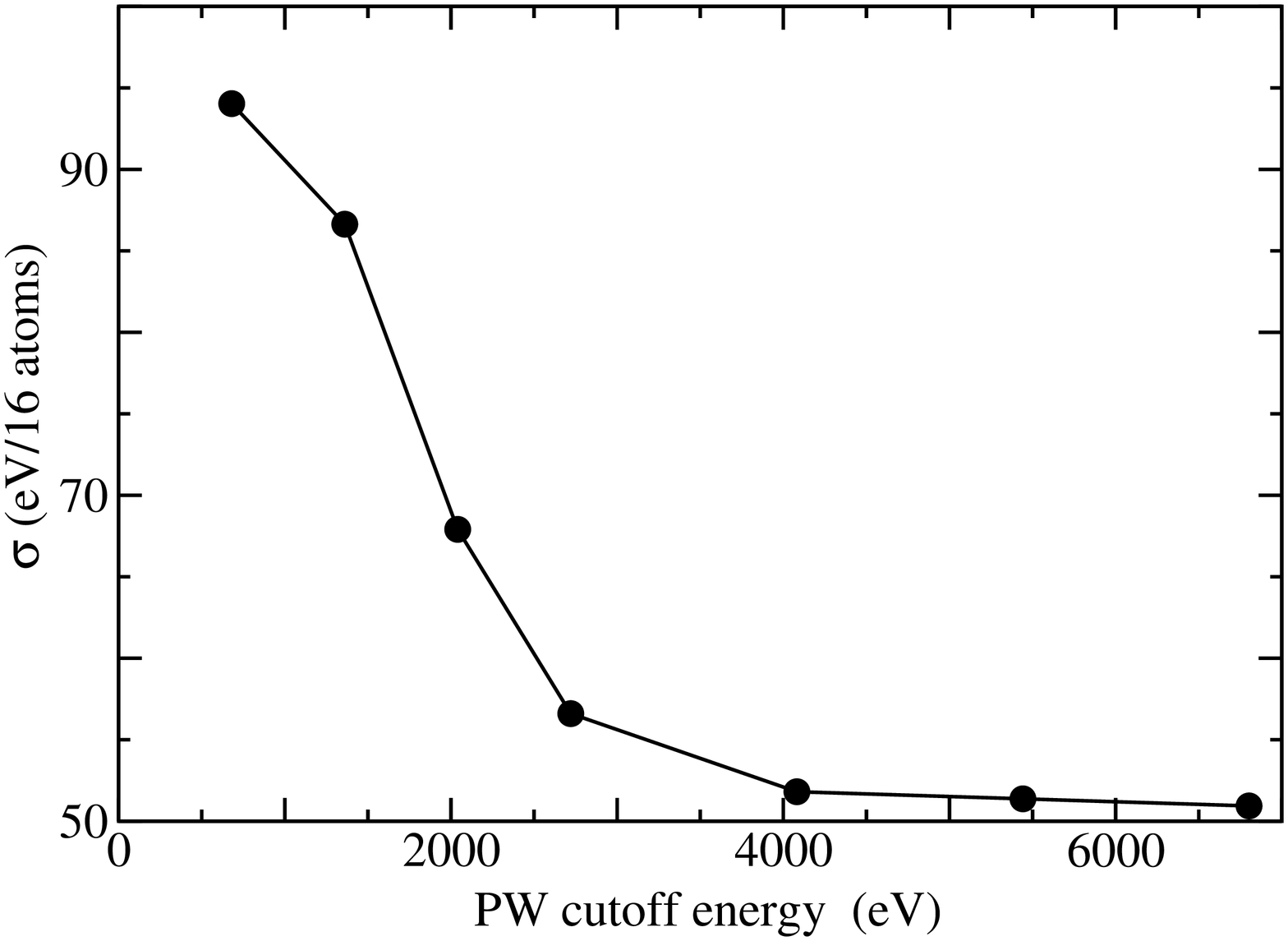,height=3.4in,angle=-90}
\caption{The standard error in the energy, $\sigma$, for runs of the
same length, calculated within VMC without a Jastrow factor, as a
function of the plane-wave cut-off energy, for a 16-atom cell of MgO
in the rock-salt structure with a lattice parameter of
4.17~\AA.}\label{fig:cutoff}
\end{figure}

We have found that the quality of our Slater-Jastrow trial wave
function is improved if a large plane-wave cut-off is used in
generating the single-particle orbitals, and a correspondingly small
blip-grid spacing is used in representing them. In order to
investigate this question, we performed a series of VMC calculations,
and calculated the standard error in the energy as a function of
plane-wave cut-off. The blip-grid spacing was taken to be related to
the plane-wave cut-off by the `natural' formula mentioned in
Sec. 2. The calculations were performed without a Jastrow factor,
because this makes it possible to check some components of the total
energy against DFT calculations.  These calculations were performed on
a 16-atom cell for MgO in the rock-salt structure with a lattice
parameter of $a=4.17$~\AA, which is close to the zero pressure
equilibrium lattice parameter.  Results are presented in
Fig.~\ref{fig:cutoff}. We notice that by increasing the PW cutoff from
680~eV to 6800~eV the standard error in the energy is reduced by a
factor of $\sim 2$.  This means that QMC runs performed using the
trial wave function obtained with the largest cutoff can be 4 times
shorter, in order to achieve the same statistical accuracy. More
importantly, we found that using a very large PW cutoff was essential
for having stable DMC runs. We were unable to perform any useful DMC
simulation with cutoff energies less than 2712~eV.

\begin{figure}
\psfig{figure=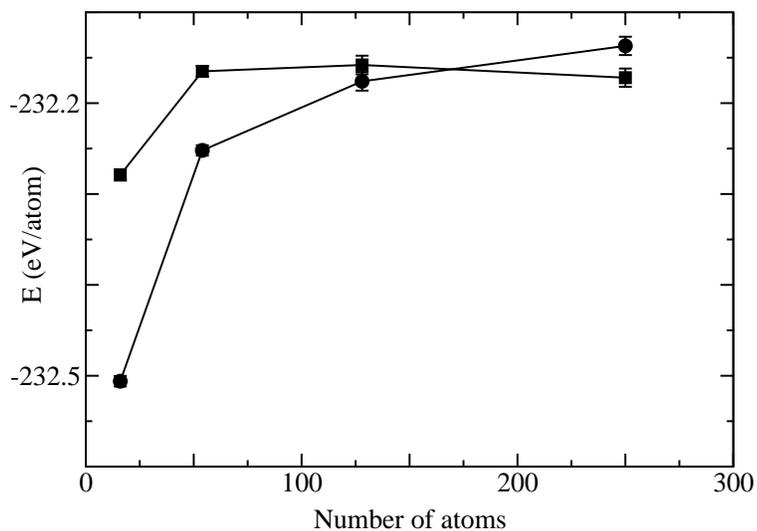,height=3.4in,angle=-90}
\caption{The VMC energy per atom for MgO in the rock-salt structure
with a volume per atom of 9.06~\AA$^3$ as a function of the number of
atoms in the repeating cell, using both the Ewald interaction (circle)
and the MPC interaction (squares).}\label{fig:size1}
\end{figure}

We have made extensive tests on system size effects. We divide our
discussion of these tests into two parts: first, tests on the
rock-salt structure at low pressures, which are relevant to the
equilibrium properties of this structure; second, tests on both the
rock-salt and CsCl structures at high pressures, which are relevant to
the determination of the transition pressure. As we shall see, these
two sets of tests involve somewhat different questions.  In
Fig.~\ref{fig:size1}, we show the VMC energy per atom of MgO as a
function of the number of atoms in the repeating cell, using both the
standard Ewald interaction and the MPC interaction. For these
calculations we used a plane-wave cut-off of 6800 eV and the
associated natural blip-grid spacing for the description of the
single-particle orbitals. We note that the MPC results appear to
converge considerably faster than the Ewald ones, and that for a
system of 54 atoms the MPC energy is already converged to better than
$\sim$ 50 meV/atom. We therefore decided to use this cell size to
evaluate the energy-volume curve presented in the following section.

\begin{figure}
\psfig{figure=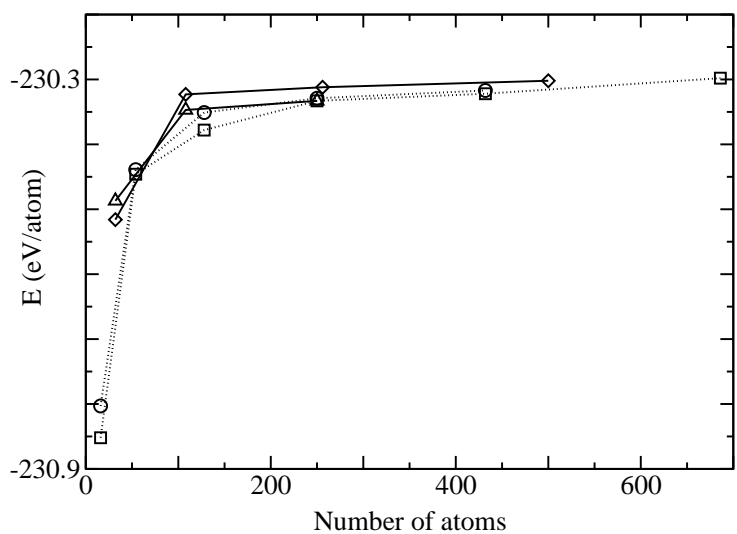,height=3.4in,angle=-90}
\caption{The VMC energy per atom for MgO in the CsCl structure with a
volume per atom of 8.77~\AA$^3$ as a function of the number of atoms
in the repeating cell, calculated using the Ewald interaction.
Squares and circles: simple cubic cell with $\Gamma$-point 
and zone-boundary
sampling, respectively; diamonds and triangles: fcc cell with 
$\Gamma$-point and zone-boundary sampling (see text for
wavevector used in zone-boundary sampling).
The lines are guides to the eye.
}\label{fig:size2}
\end{figure}

The results we report here for the CsCl structure were performed with
the standard Ewald method rather than the MPC method.  In order to
check system-size errors thoroughly, we found it essential to perform
tests on large systems of up to over 600 atoms. We made extensive
tests on the CsCl structure to compare sc and fcc repeating geometries
and to examine the effect of using different sampling wavevectors.
The sampling wavevectors we used are the $\Gamma$-point
$( 2 \pi / L ) (0,0,0)$, and the wavevector $( 2 \pi / L ) (0.5,0.5,0.5)$,
where $L$ specifies the dimension of the repeated cell. (In more
detail, $L$ is the length such that with sc repeating geometry
the primitive translation vectors are $L (1,0,0)$, $L (0,1,0)$
and $L (0,0,1)$, while with fcc geometry they are
$L (0,0.5,0.5)$, $L (0.5,0,0.5)$ and $L (0.5,0.5,0)$.)
Since the wavevector $( 2 \pi / L ) (0.5,0.5,0.5)$ lies on the
zone boundary of the Brillouin zone associated with the periodically
repeated supercell, we refer to sampling using this wavevector
as ``zone-boundary'' sampling.
These tests were performed
at the volume 8.77~\AA$^3$/atom, which is close to the zero pressure
equilibrium volume. The tests were all performed using VMC, and we
used the Jastrow factor optimised using a 16 atom cell for all system
sizes, because re-optimising the Jastrow factor introduces small
``jumps'' in the energy.  Since we needed to go to large system sizes,
we decided to reduce the plane-wave cut-off from 6800 eV to 2712 eV,
because this gave a considerable reduction in the memory required;
with this lower cut-off, the standard error in the energy fluctuations
is only slightly larger (see Fig.~\ref{fig:cutoff}), and DMC
calculations are still stable. Results of these tests are shown in
Fig.~\ref{fig:size2}. We see that convergence to within less than 50
meV/atom is obtained for systems larger than 108/128 atom.  We also
note that convergence is better with the fcc than with the sc
repeating geometry, as expected, and that there is little to
choose between $\Gamma$-point and ``zone-boundary'' sampling.
We have therefore performed all further calculations using
fcc geometry and $\Gamma$-point sampling.

Since calculation of the transition pressure requires QMC calculations
for the two structures at high pressures, we have performed further
VMC calculations at volumes 4.23 and 4.41~\AA$^3$/atom for the CsCl
and the rock-salt structures respectively, close to the transition, using
the Ewald interaction. Results of these tests are reported in
Table~\ref{tab:size}. We see that for calculations on the 
CsCl structure, using a
cell containing 108 atoms with fcc repeating geometry, 
the error is about 50~meV/atom. 
The error is approximately the same for the rock-salt structure 
with a 128-atom
cell, so that the error in the energy difference between the two
structures is less than our target accuracy of 30 eV/atom.

\subsection{Production results}

\begin{figure}
\psfig{figure=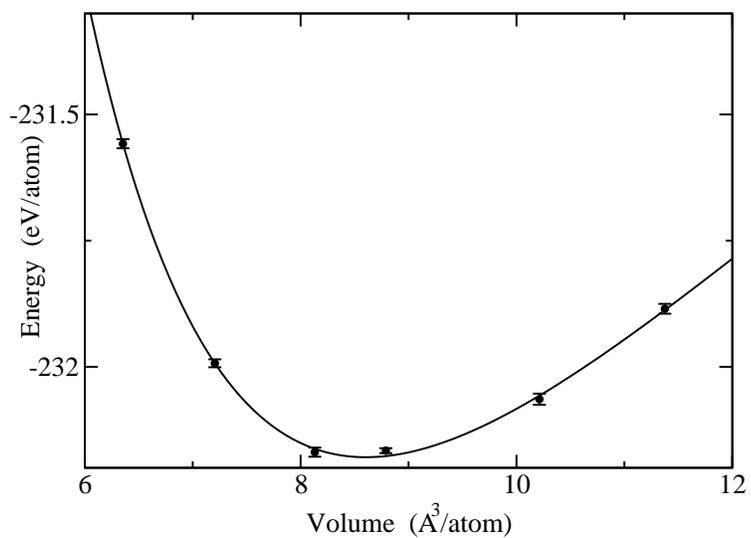,height=3.4in,angle=-90}
\caption{The DMC energy per atom as function of volume for MgO in the
NaCl structure.}\label{fig:nacl}
\end{figure}

In Fig.~\ref{fig:nacl} we display DMC energies as a function of volume
for MgO in the NaCl structure. The length of these simulations was
typically 6000 steps, resulting in a statistical error bar of less
than 10 meV/atom.  These energy points were then used to fit the
parameters of the Birch-Murnaghan equation of state~\cite{birch47}:
\begin{equation}\label{murna}
E = E_0 + \frac{3}{2}V_0B_0 \left [ \frac{3}{4}(1+2\xi)\left
(\frac{V_0}{V}\right )^{4/3} - \frac{\xi}{2} \left ( \frac{V_0}{V}
\right )^{2} -\frac{3}{2}(1+\xi) \left ( \frac{V_0}{V} \right )^{2/3}
+ \frac{1}{2} \left ( \xi + \frac{3}{2}\right ) \right ] \; ,
\end{equation} 
where $\xi = (3 - 3 B_0'/4)$, $V_0$ is the equilibrium volume, $B_0$
is the zero-pressure bulk modulus, $B_0'$ is its derivative with
respect to pressure at zero pressure, and $E_0$ is the energy at the
minimum.  The fitted curve is also reported on the same Figure. The
values of the fitted parameters are reported in
Table~\ref{tab:murna_parameters} together with other theoretical
results and experimental data.  A comparison of the QMC results with
experimental values shows that our calculated lattice parameter of
$a_0=4.098$~\AA\ is smaller than the measured value of
$a_0=4.213$~\AA\ \cite{fei99}, and our bulk modulus $B_0=183$~GPa is
greater than the measured value $B_0=160 \pm
2$~GPa~\cite{fei99}. However, two kinds of corrections need to be
made. It is known from earlier DFT calculations~\cite{karki00} that
room temperature thermal pressure due to lattice vibrations increases
$a_0$ by $0.03$~\AA\ and decreases $B_0$ by 10 GPa. We should also
correct for pseudopotential errors. To estimate these, we have
compared the predictions of pseudopotential and all-electron HF
calculations using the {\sc crystal}~\cite{crystal} code (see Table
1). This shows that the pseudopotentials we have used underestimate
$a_0$ by 0.10~\AA\ and overestimate $B_0$ by 15 GPa. Combining these
two corrections, our revised QMC values are $a_0=4.23$~\AA\ and
$B_0=158$~GPa, which are very close to the experimental values.

\begin{figure}
\psfig{figure=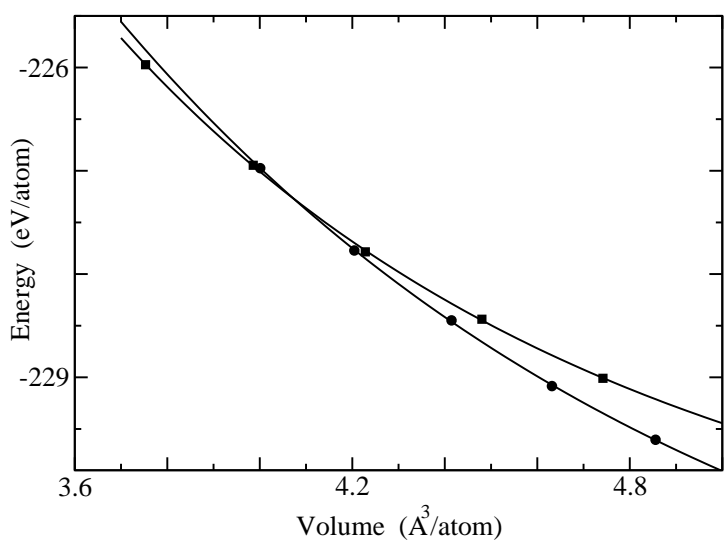,height=3.4in,angle=-90}
\caption{The DMC energy per atom for the NaCl structure (circles) and
the CsCl structure (squares) of MgO as a function of volume in the
region of the B1-B2 phase transition.}\label{fig:transition}
\end{figure}

\begin{figure}
\psfig{figure=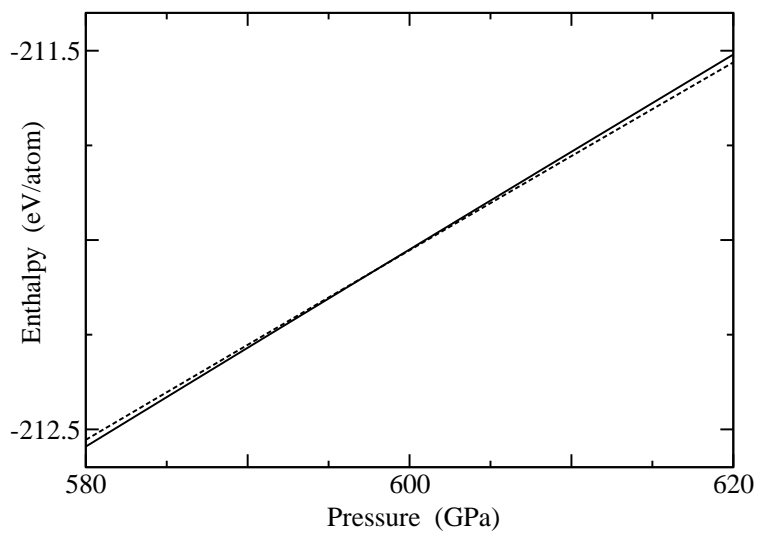,height=3.4in,angle=-90}
\psfig{figure=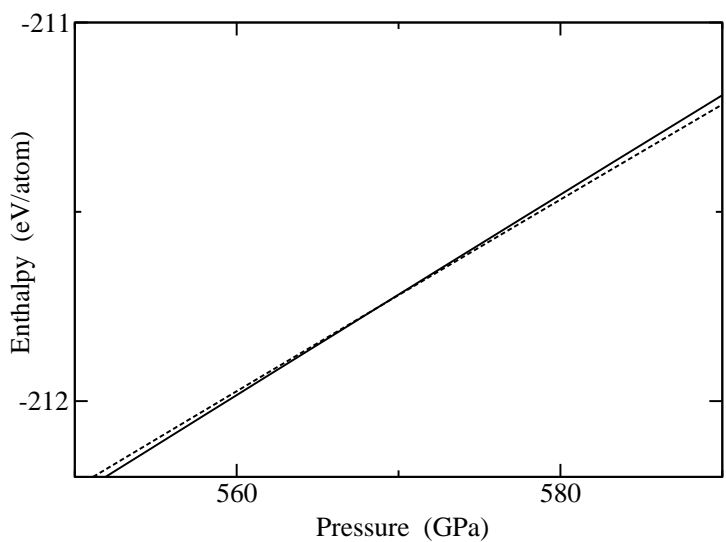,height=3.4in,angle=-90}
\caption{The enthalpy per atom of the NaCl and CsCl (dashed line)
structures of MgO as a function of pressure. Top picture: fit to the
DMC data; bottom picture: fit to the DFT-LDA data.}\label{fig:enth}
\end{figure}

In Fig.~\ref{fig:transition} we report the DMC energy for MgO in the
NaCl and the CsCl structures evaluated for volumes corresponding to
roughly one half of the zero pressure equilibrium volume, which is the
region of volumes in which the transition occurs. These energy points
have also been fitted to the Birch-Murnaghan equation of state, which
we have then used to compute the enthalpies of the two structures, which
are displayed in the upper part of Fig.~\ref{fig:enth}, from which we
infer a transition pressure of about 597 GPa. We note that the slopes
of the two curves are very similar, and that an error of about 1
meV/atom in the relative enthalpy results in an error in the
transition pressure of about 1 GPa. We do not expect our DMC results
to be more accurate than about 20 meV/atom, so our computed transition
pressure should be considered to have an error bar of about 20 GPa.
For comparison, we also report in the lower part of
Fig.~\ref{fig:enth} the enthalpy evaluated with DFT-LDA and the same
pseudopotentials, from which we deduce a transition pressure of 569
GPa.

\section{Discussion and conclusions}

An important conclusion from this work is that it is technically
feasible to carry out diffusion Monte Carlo calculations on MgO at the
level of accuracy required to compute quantities such as the
equilibrium lattice constant, the bulk modulus, and the B1-B2
transition pressure. As we mentioned in the Introduction, this is a
non-trivial conclusion, because DMC calculations succeed in practice
only if the many-body trial wave function is sufficiently close to the
true ground state wave function. Even though MgO should be a
favourable oxide in this respect, it was still essential to pay
careful attention to the accurate representation of the
single-electron orbitals in order to bring statistical fluctuations
under control.

We have shown that, provided corrections are made for thermal effects
and errors due to imperfections of our pseudopotentials, the QMC
predictions of lattice parameter $a_0$ and bulk modulus $B_0$ agree
with experimental values to within $\sim 0.5$\% for $a_0$ and to
within experimental error ($\sim 2$\%) for $B_0$. In future work,
there should be scope for further improvement in the
pseudopotentials. It is interesting to note that our QMC prediction
for $a_0$ is almost identical to the HF prediction with the same
pseudopotentials. This might seem to suggest that correlation effects
are negligible in MgO. However, this is certainly not the case. The
correlation energy is at least the difference between the HF and the
QMC total energies. We find that for the rock-salt structure near the
equilibrium volume, this difference is $\sim 4.5$~eV/atom. The close
agreement between the HF and QMC lattice parameters therefore
indicates that this rather large correlation energy depends only
weakly on volume. Our QMC value of $597 \pm 20$~GPa for the B1-B2
transition pressure supports the most recent DFT predictions of a
pressure in the region of $\sim 500$~GPa, which is beyond the region
of geophysical interest (the pressure at the core-mantle boundary of
the Earth is 135 GPa). The detailed value we have found may suffer
from a pseudopotential error, but at present we are unable to quantify
this.

With these encouraging results for MgO, we believe that there are now
good prospects for extending QMC methods to studying the more
challenging problems involving MgO mentioned in the introduction, and
to studies of transition metal oxides. LDA (and generalised gradient
approximation) calculations are unsuitable for transition metal oxides
because they lead to an incorrect filling of the energy levels.
Unrestricted HF and B3LYP orbitals have already been used with some
success in DMC studies of transition metal
oxides~\cite{needs_2003,lee_2004}, and one might also consider using
LDA+U or SIC (self-interaction corrected) DFT orbitals.  Towler and
Needs~\cite{towler_2005} found that unrestricted HF orbitals gave a
lower energy than B3LYP orbitals for NiO.  Transition metal oxides are
clearly a case where it would be useful to optimise the orbitals in
the presence of the Jastrow factor.


\begin{table}
\begin{tabular}{lccc}
\hline
          & $a_0$ (\AA) & $B_0$ (GPa) & $P_{\rm tr}$(B1-B2) (GPa)\\ 
\hline
Experiments & 4.213$^a$  4.211$^b$ 4.212$^c$ 4.19$^d$ &  160$\pm 2^a$ 160.2$^c$ 156$^e$ 164.6$^d$ & $>$ 227$^f$   \\
QMC & 4.098$^g$               & 183$^g$     &   597$\pm 20^g$  \\
HF-PP & 4.089$^g$               & 196$^g$     &     \\
HF-AE & 4.195$^{g,h}$ 4.200$^i$  & 181$^{g,h}$ 182$^i$ &    \\
HF-LCAO & 4.201$^j$ 4.191$^q$    & 186$^j$ 182$^q$    & 220$^j$ 712$^q$  \\
B3LYP & 4.230$^{i}$     & 162$^{i}$     &    \\
DFT-LDA & 4.160$^i$ 4.240$^k$ 4.194(4.222)$^l$ & 198$^i$ 172.6$^k$ 169(159)$^l$ &  490$^k$ 451$^m$ 510$^n$   \\
  &  4.25$^m$ 4.167$^n$ 4.163$^o$ 4.191$^p$ 4.160$^q$ & 159.7$^m$ 172$^n$ 185.9$^o$ 146$^p$ 181$^q$& 515$^o$ 1050$^p$ 512$^q$\\
DFT-GGA & 4.273$^q$ 4.243$^q$ 4.247$^o$ 4.244$^i$ & 153$^q$ 160$^q$ 159$^q$ 169.1$^o$ 157$^i$ & 478$^q$ 428$^q$ 418$^q$ 515$^o$  \\
 &  4.253$^r$ 4.259$^s$ 4.218$^u$ 4.259$^v$ & 150.6$^r$ 161.5$^u$ 160$^v$ & 509$^r$ 664$^s$ 400$^u$ \\
\hline
\end{tabular}
$^a$ Ref.~\protect\cite{fei99}\\
$^b$ Ref.~\protect\cite{wyckoff63}\\
$^c$ Ref.~\protect\cite{speziale01}\\
$^d$ Ref.~\protect\cite{anderson66}\\
$^e$ Ref.~\protect\cite{mao79}\\
$^f$ Ref.~\protect\cite{duffy95}\\
$^g$ This work.\\
$^h$ Ref.~\protect\cite{mccarthy94}\\
$^i$ Ref.~\protect\cite{marinelli03}\\
$^j$ Ref.~\protect\cite{causa86}\\
$^k$ Ref.~\protect\cite{oganov03}\\
$^l$ Ref.~\protect\cite{karki00}; values in parenthesis include
zero point motion and room temperature effects.\\
$^m$ Ref.~\protect\cite{karki97}\\
$^n$ Ref.~\protect\cite{mehl88}\\
$^o$ Ref.~\protect\cite{jaffe00}\\
$^p$ Ref.~\protect\cite{chang84}\\
$^q$ Ref.~\protect\cite{habas98}\\
$^r$ Ref.~\protect\cite{oganov03prb}\\
$^s$ Ref.~\protect\cite{drummond02}\\
$^u$ Ref.~\protect\cite{strachan99}\\
$^v$ Ref.~\protect\cite{tsuchiya01}\\
\caption{DFT and HF values for the lattice constant and bulk modulus
of the NaCl phase of MgO, and the equilibrium pressure for the B1-B2
transition. See the original references for
details.}\label{tab:murna_parameters}
\end{table}

\begin{table}
\begin{tabular}{ccc}
\hline
Number of atoms & \multicolumn{2}{c}{Energy (eV/atom)} \\
\hline
  &   NaCl          &    CsCl     \\
\hline
32   &              & -227.143(6) \\
54   & -227.971(4)  &             \\
108  &              & -226.914(4) \\
128  & -227.846(3)  &             \\
250  & -227.806(5)  &             \\
256  &              & -226.867(5) \\
432  & -227.794(7)  &             \\
500  &              & -226.866 (9) \\
\hline
\end{tabular}
\caption{VMC energies for MgO in the NaCl and CsCl structures at
volumes per atom 4.41 and 4.23~\AA$^3$ respectively, as a function of
the number of atoms in the repeating cell.}\label{tab:size}
\end{table}

\end{document}